# Effects of Substrate Heating and Wettability on Evaporation Dynamics and Deposition Patterns for a Sessile Water Droplet Containing Colloidal Particles

Nagesh D. Patil, Prathamesh G. Bange, Rajneesh Bhardwaj[*], Atul Sharma

Department of Mechanical Engineering,

Indian Institute of Technology Bombay, Mumbai, 400076 India

[*]Corresponding author (rajneesh.bhardwaj@iitb.ac.in)

Phone: +91 22 2576 7534, Fax: +91 22 2572 6875

*Abstract*

Effects of substrate temperature, substrate wettability and particles concentration are experimentally investigated for evaporation of a sessile water droplet containing colloidal particles. Time-varying droplet shapes and temperature of the liquid-gas interface are measured using high-speed visualization and infrared thermography, respectively. The motion of the particles inside the evaporating droplet is qualitatively visualized by an optical microscope and profile of final particle deposit is measured by an optical profilometer. On a non-heated hydrophilic substrate, a ring-like deposit forms after the evaporation, as reported extensively in the literature; while on a heated hydrophilic substrate, a thinner ring with an inner deposit is reported in the present work. The latter is attributed to Marangoni convection and recorded motion of the particles as well as measured temperature gradient across the liquid-gas interface confirms this hypothesis. The thinning of the ring scales with the substrate temperature and is reasoned to stronger Marangoni convection at larger substrate temperature. In case of a non-heated hydrophobic substrate, an inner deposit forms due to very early depinning of the contact line. On the other hand, in case of a heated hydrophobic substrate, the substrate heating as well as larger particle concentration helps in the pinning of the contact line, which results in a thin ring with an inner deposit. We propose a regime map for predicting three types of deposits namely, ring, thin ring with inner deposit and inner deposit - for varying substrate temperature, substrate wettability and particles concentration. A first-order model corroborates the liquid-gas interface temperature measurements and variation in the measured ring profile with the substrate temperature.

# 1 Introduction

Evaporation of a sessile water droplet containing colloidal particles on a substrate is a much-studied problem for the last two decades. Potential technical applications include inkjet printing [1], surface coatings[2], cooling of heated surfaces[3,4] and biosensors[5,6]. The physics involved during the droplet evaporation is a complex interplay of several multi-scale transport phenomena, namely, heat conduction and convection in the droplet, heat conduction in the substrate, liquid-vapor diffusion outside the droplet, advection of the colloidal particles in the droplet and moving contact line during the evaporation (see recent review by Larson[7]). The evaporation dynamics of the droplet depends on the motion of the contact line and in general, the evaporation is a two-stage process[8], namely, constant contact radius (CCR) mode and constant contact angle (CCA) mode. In the CCR mode, the droplet contact line is pinned to the surface and contact angle decreases with time; while in the CCA mode, the contact angle remains constant and wetted radius decreases with time due to the depinning of the contact line.

As established by Deegan et al.[9], the evaporation mass flux along the liquid-gas interface is non-uniform and is the largest near the contact line, which creates radially outward fluid flow inside the droplet. Thus, an evaporating droplet dispersed with colloidal particles results in a ring-like particle deposit, due to the advection of the particles by the outward fluid flow, a phenomenon named as "coffee-ring" effect[9]. Due to the non-uniform evaporation mass flux, the liquid-gas interface cools non-uniformly driven by latent heat of evaporation. Owing to the temperature gradient on the interface, a thermal Marangoni convection loop develops in the droplet[10]. The deposit pattern is influenced by the Marangoni convection since particles are advected back from the contact line to the droplet center. Therefore, the deposit is an accumulation of the particles at the center of the droplet instead of the ring[10,11].

Owing to the technical applications, several studies have investigated ways to control the spatial distribution of the particles deposit. For instance, in inkjet printing, a uniform deposit is desired than the ring-like deposit. In addition to the Marangoni convection, other effects which were found important in controlling the shape of the deposit are pH of the colloidal suspension[12], particles hydrophobicity[13], shape of the particles[14, 15], surfactant concentration[16], substrate elasticity[17], ambient pressure[18], particles size[19] and bio-molecular interaction between the particles and substrate[20]. The present study investigates the influence of substrate temperature, substrate wettability and particle concentration on the mechanism of the particles deposition and associated evaporation dynamics. In the following sub-sections, previous studies which investigated these effects are described.

## 1.1 Studies on non-heated, hydrophilic substrates

The Marangoni convection has been found important to control the particle deposition on non-heated, hydrophillic substrates in several previous studies. Hu and Larson[10] reported the suppression of ring-like deposit for microliter octane droplets evaporating on glass coated with perfluorolauric acid. Ristenpart et al.[21] showed that the direction of the Marangoni convection depends on the substrate-droplet thermal conductivity ratio and consequently, it influences the final deposit pattern. They considered droplets of isopropanol, chloroform, methanol and ethanol on polydimethylsiloxane (PDMS) substrate. Bhardwaj et al.[11] showed that the Marangoni convection advects all the colloidal particles in an evaporating nanoliter isopropanol droplet on a PDMS substrate, to form a central bump pattern. Xu et al.[22] showed that the thickness of substrate influences the temperature gradient on the liquid-gas interface and direction of Marangoni convection. Recently, Weon and Je[23] showed that competition between the radially outward flow and Marangoni flow induces a fingering like pattern in a decalin droplet containing micro- and nanoparticles particles. Yang et al.[24] used water droplets containing sulfate-modified polystyrene particles with different concentrations (0.1 to 0.5% v/v) and obtained various patterns (concentric multi-rings, radial spokes, spider web, foam and island like depositions) as a result of the opposition between the receding contact line and growth rate of the particle deposition during the evaporation.

**1.2  Studies on non-heated, hydrophobic substrates**

Regarding evaporation of the droplets on non-heated, hydrophobic substrates, Orejon et al.[25] showed that $TiO_2$ nanoparticles promote the pinning of the contact line as well as stick-slip behavior. They reported that a larger concentration of the particles delays the depinning of the contact line as compared to that for respective pure liquid droplet. Recently, Nguyen et al.[26] obtained centralized deposits on a silanized hydrophobic silicon substrate, due to the depinning of contact line which brings silica particles to inner region of the droplet. Further, Li et al.[27] controlled polymer nanoparticles deposition by varying the contact angle hysteresis (CAH) of the substrates. They reported centralized deposits for weak CAH substrate (such as sodium polysulfonate) and ring patterns on strong CAH substrates (such as polyvinyl pyrrolidone).

**1.3  Studies on heated, hydrophilic substrates**

In the context of heated, hydrophillic substrates, more recently Li et al.[28] studied the effect of heating of glass substrate from $T_s$ = 30 to 80°C on the final deposit patterns. The equilibrium contact angles ($\theta_{eq}$) of water droplets containing 0.25 % (v/v) polystyrene nanoparticles on the substrate were in range of 24 - 30°. The pattern is a thin ring with an inner deposit (named as "coffee eye") on the heated substrate. Similarly, Parsa et al.[29] varied hydrophilic silicon substrate temperature ($\theta_{eq}$ = 30 - 40°, $T_s$ = 25 - 99°C) and obtained dual-ring patterns for water droplet containing 0.05 % (w/w) copper oxide nanoparticles. The thinning of the ring in these two studies [28, 29] was attributed to Marangoni velocity which transports the particles from the contact line region to the droplet inner region.

The Marangoni velocity scales with the temperature gradient along the liquid-gas interface [12] and previous studies reported temperature measurements of the interface using infra-red thermography for pure liquid droplets[30, 31]. The direct measurement of the Marangoni velocity inside an evaporating pure liquid droplet was reported by Savino and Fico[30] by recording motion of tracers using laser sheet illumination. More recently, Pradhan and Panigrahi[32] utilized a confocal, microscale PIV technique to measure the Marangoni velocity in an evaporating pure liquid droplet. We note that the PIV with tracers is an intrusive measurement technique in case of a droplet with colloidal particles.

### 1.4 Objectives of the present study

Previous studies[28, 29] in the context of the *heated*, *hydrophilic* substrates investigated the effect of substrate temperature at a constant particles concentration, however, coupled effects of the particle concentration and substrate temperature on the deposition patterns have not been reported thus far. In addition, to the best of our knowledge, there is no study available which investigates the evaporation of droplets containing colloidal particles on the *heated*, *hydrophobic* substrate and previous reports considered only *non-heated*, *hydrophobic* substrates[25, 26, 27]. With respect to the effect of particle concentration, Orejon et al.[25] showed that the particles concentration influences the depinning of the contact line on *non-heated hydrophobic* substrate, however, the effect of the substrate heating as well as coupled effects of substrate heating and particle concentration on the depinning are unknown. The focus of the present work is to understand above-mentioned effects in order to control the particles deposition pattern. We consider evaporation of microliter water droplets containing 0.46 μm polystyrene particles and systematically vary substrate wettability (hydrophilic glass and hydrophobic silicon), substrate temperature ($T_s$ = 27.5°C, 60°C and 90°C) and particles concentration ($c$ = 0.05%, 0.1% and 1.0%, v/v). We combine the following measurement techniques in the present work:

- High-speed visualization to record the temporal variation of volume, wetted diameter, contact angle and height of the droplet (section 2.2),
- Infrared thermography to measure the spatial and temporal distribution of the liquid-gas interface temperature (section 2.3),
- Optical microscopy to record the particle motion in the evaporating droplet and capture the final deposit image, and optical profilometer to measure the ring-like deposit (section 2.4),

The effects of substrate temperature (section 3.2), substrate wettability (section 3.3) and particles concentration (section 3.4) on the evaporation dynamics and deposit shapes are presented. In section 3.5, first-order models are formulated to verify the temperature measurements and mass of the particles deposited in the ring. Finally, a regime map for predicting the deposit shape as function of substrate temperature, substrate wettability and particle concentration is proposed in section 3.6.

## 2 Methods

### 2.1 Preparation of colloidal solution

An aqueous colloidal suspension, with 10% volume concentration (v/v) of uniformly dispersed polystyrene latex beads of diameter ~ 0.46 μm, was obtained from Sigma Aldrich Corporation, USA (LB5, particle density ~ 1005 kg/m³). Using this solution, we prepared the solutions of particles concentration, $c$ = 0.05, 0.1 and 1% v/v, by diluting it with deionized (DI) water. The prepared solutions were stabilized with sufficient sonication (about 1 hour). It was ensured that no sedimentation or agglomeration of particles occurred in enclosed microcentrifuge tubes (Eppendorf Inc.), which were used for the storage of the prepared solutions.

## 2.2  Droplet generation and high-speed visualization

Deionized water droplets of 1.1±0.2 µL volume were generated using a micropipette (Prime, Biosystem Diagnostics Inc., India) having fixed volume dispensing capacity. The droplets were gently deposited on hydrophilic glass and hydrophobic silicon substrates. Glass slides (75×15×1.2 mm) were sequentially cleaned by isopropanol and deionized water, and were allowed to dry completely in ambient. Silicon wafer (diameter = 51 mm, thickness = 0.3 mm) with polished side was cleaned by standard RCA cleaning process and wet oxidized in a furnace.

The evaporation of the sessile droplet on the substrate was visualized from the side as shown by a schematic in Figure 1. The visualization was performed using a high-speed camera (IDT Inc, MotionPro, Y-3 classic) with a long distance working objective (Qioptiq Inc.), similar to setup in our recent work[33]. A white LED lamp was used as a back light source for the high-speed visualization. The images were recorded at 10 and 100 frames per second (fps) for non-heated and heated substrates, respectively. The working distance, image resolution and pixel resolution are 9.5 cm, 600 × 450 and 14 µm, respectively. The data of the recorded images are analyzed using MATLAB® image processing modules[34]. Since the droplet wetted diameter is lesser than the capillary length of water (~2.7 mm), the droplet assumes a spherical cap during the evaporation [35]. Thus, the droplet volume $v$ and the contact angle $\theta$ are calculated using the following equations,

$$v = (1/6)\pi h_L [3(d_{w,i}/2)^2 + h_{L,i}^2] \text{ and } \theta = 2\tan^{-1}[h_{L,i}/(d_{w,i}/2)] \qquad (1)$$

where $d_{w,i}$ is the wetted diameter and $h_{L,i}$ is height of the droplet, as shown in second row of Figure 2a. The equilibrium contact angle ($\theta_{eq}$) was calculated using above equation, and advancing ($\theta_{adv}$) and receding ($\theta_{rec}$) contact angles were measured using tilting plate method. $\theta_{eq}$, $\theta_{adv}$, $\theta_{rec}$ and contact angle hysteresis ($\theta_{adv}$ - $\theta_{rec}$) are listed in Table 1 for non-heated as well as heated glass and silicon substrates. The overall uncertainty in measurements for contact angle and wetted diameter/height are around ±3-4° and ±28 μm, respectively. Each experiment was performed three times to ensure repeatability. The ambient temperature and relative humidity were 27.5±1.5°C and 35±4%, respectively, for the present measurements.

## 2.3 Temperature measurement using infrared thermography

The substrates were heated to different temperatures ($T_s$ = 60 and 90°C) using a hot plate of 200 mm diameter by a controlled power supply. The droplets were deposited on the substrates after its surface attained a desired temperature and remained steady for at least 4-5 minutes. From the top (Figure 1), spatial as well as temporal temperature distribution of the liquid-gas and solid-gas interface is recorded using an infrared (IR) camera (A6703sc, FLIR Systems Inc.), with 25 mm (f/2.5) IR lens and a close focusing extender ring. Since water is generally opaque to the infrared radiation, the measured temperature is essentially the liquid-gas interface temperature (referred as *droplet surface temperature*). The imaging rates were kept same as that for the high-speed visualization *i.e.* 10 and 100 frames per second (fps) for non-heated and heated substrates, respectively. The working distance, image resolution and pixel resolution are 10.6 cm, 300 × 256, and 16 μm per pixel, respectively. The values of emissivity for water, glass, and silicon wafer were taken as 0.97, 0.95, and 0.78, respectively. The uncertainty in the temperature response of the infrared camera is around ±0.6°C and calibration curve is given in the supplementary information.

## 2.4 Visualization of motion of particles and measurement of particle deposits

The colloidal particles were visualized from the top by a high-speed camera (IDT Inc., MotionPro, Y-3 classic) mounted on an optical microscope (Olympus MX-40, objective lens 2.5X to 50X).

The deposit profiles were quantitatively measured by a 3D optical profilometer (Zeta-20, Zeta Instruments Inc.). The measured ring profiles at four different azimuthal locations were averaged to obtain the final ring profile, as described in the supplementary information.

## 3  Results and discussions

In this section, the results for the evaporation of water droplets on *two substrates of different wettabilities* (hydrophilic glass and hydrophobic silicon) maintained at *three temperatures* ($T_s$ = 27.5, 60 and 90°C) for *three particles concentration* of the solution ($c$ = 0.05%, 0.1% and 1.0%, v/v) are presented. First, we present the evaporation of a water droplet on glass at ambient temperature $T_s$ = 27.5°C and $c$ = 0.05% in section 3.1. Thereafter, as compared to the ambient case, the effects of substrate heating, substrate wettability and particles concentration on the evaporation dynamics and mechanisms of the particle deposition are presented.

### 3.1  Evaporation of the droplet on glass at ambient temperature

The evaporation of a 1.1±0.2 µL water droplet on a glass substrate at ambient temperature, $T_s$ = 27.5°C and particle concentration, $c$ = 0.05% (v/v) is studied. The initial values (at $t$ = 0 s) of the wetted diameter, droplet height, contact angle are $d_{w,i}$ = 2.52 mm, $h_{L,i}$ = 0.4 mm, $\theta_i$ = 34.3°, respectively. The time-wise variation of the droplet shapes (captured using high-speed visualization) and droplet *surface isotherms* (obtained using infra-red thermography) are shown in Figure 2a and Figure 2b, respectively. A horizontal green line in Figure 2a demarcates the droplet from its reflection on the substrate. The temporal variations of *normalized* droplet wetted-diameter $D_w$, droplet height $H$, contact angle $\theta$, and volume $V$ are plotted in Figure 2c and the parameters are normalized by their respective initial values. Figure 2d shows the temporal variation of the droplet surface temperature profiles, along a horizontal line, shown in second row of Figure 2b.

During the droplet evaporation, Figure 2c shows that the wetted diameter remains constant for most of the evaporation time, while the contact angle and volume decreases monotonously *i.e.* CCR mode of the evaporation, except a sharp decrease towards the end of the process. Before the droplet deposition on the substrate, at $t = -60$ s, Figure 2d plots almost constant glass surface temperature. The figure shows that the measured droplet surface temperature near the *contact line* (radial distance, $r = 1.26$ mm) is higher than that of the *apex-point* ($r = 0$) temperature. The apex temperature in the initial stage of evaporation is smaller by around 3°C than in the later stages ($T_{apex} \approx 23.2$°C at $t = 0.1$ s and $T_{apex} \approx 26$°C at $t > 3$ s). This may be explained by the fact that the droplet is relatively cooler as compared to the ambient before the deposition due to thermal equilibration with the ambient [36]. The temperature gradient along the liquid-gas interface is weak and does not induce Marangoni flow. As explained in Refs. [37, 38], this can be attributed to unavoidable contaminants on the water-air interface which suppresses Marangoni effect. Hence, the capillary-driven *outward radial flow* (induced by the high evaporation rate at the contact line as compared to that at the apex) advects the particles near to the contact line (see Movie S1, ESI[1]). After evaporation, the final *deposition-pattern* is *ring-like* deposit, discussed later and shown in the top inset of Figure 4a for the non-heated substrate.

## 3.2 Effect of substrate heating

In order to study the effect of substrate temperature on the shape of the particles deposit, the glass substrate is heated to $T_s = 60$°C and $T_s = 90$°C and the temporal variation of the wetted diameter, height, contact angle and volume is plotted in Figure 3. Figure 3a and 3b show that the evaporation occurs in CCR mode for both cases of the substrate heating, similar to that discussed for the substrate at ambient temperature in section 3.1. In Figure 3c and 3d, we observe that the measured surface temperature at the contact line is much larger than that of the apex-point temperature; $\Delta T = T_s - T_{apex}$ = 4°C, 20°C, and 38°C (at $t$ = 0.1s) for the substrate temperatures of 27.5°C, 60°C and 90°C, respectively. The larger $\Delta T$ for the heated substrate generates the thermal Marangoni flow from contact line region towards the droplet apex-point, which results in a flow circulation, shown schematically in inset of Figure 3c. The particles advect by the circulation and move towards the apex. During the evaporation, it is noted that initially particles move toward the contact line region before reversing its direction due to the Marangoni effect, which qualitatively confirm the existence of two opposite flows separated by a *stagnation region* in the vicinity of contact line (see Movie S2, ESI[2]), shown later schematically in Figure 5c. The existence of the stagnation region was also reported by Li et al. [28]. Finally, at the end of the evaporation, the dried *deposition-pattern* of the particles is a *thin ring with an inner-deposit* (accumulated inside the ring); can be seen at the top of Figure 4a for the heated substrate.

---

[1]Electronic supplementary information (ESI), Movie S1 – Radial outward flow of particles inside the droplet, during evaporation on glass at ambient temperature.

The measured ring profiles at different substrate temperatures ($T_s$) are compared in Figure 4a and the images of the final deposit are given in the top inset. With increasing glass substrate temperature $T_s$, Figure 4a shows a monotonic decrease in width and height of the ring and the variation is reasoned as follows. The Marangoni velocity ($V_{Ma}$) scales with the temperature gradient across the interface [12, 38] and is expressed as,

$$V_{Ma} \sim \frac{1}{32} \frac{\beta \theta^2 \Delta T}{\mu} \qquad (2)$$

---

[2]ESI, Movie S2 – Marangoni flow inside the droplet evaporating on glass kept at 60°C temperature.

where, $\beta = d\gamma/dT$ is the gradient of surface tension with respect to the temperature (-1.68×10$^{-4}$ N/m-K), $\theta$ is the *instantaneous* contact angle, $\Delta T$ is the *instantaneous* temperature difference between the contact line and the apex-point of the droplet, and $\mu$ is the dynamic viscosity (1×10$^{-3}$ N-s/m$^2$). At four time instances (see Table 2), Figure 4b shows a monotonic increase in the $V_{Ma}$ with increasing $T_s$; thus, larger Marangoni flow brings more particles to the droplet apex-point, and decreases the width as well as height of the deposited ring. The variation in the mass of the particles in the ring with the substrate temperature is compared with predictions of a first-order model in section 3.5.

The mechanisms of the formation of the particles deposits obtained in case of non-heated and heated glass substrates are summarized schematically in Figure 5(a, c). Figure 5(b, d) shows the images of the dried deposits in these cases. For the *non-heated* substrates, Figure 5a shows a qualitative variation in the evaporative mass flux on the liquid-gas interface, with the largest evaporation flux near the contact line [9, 39], resulting in the radially outward flow with pinned contact line. The flow advects almost all particles from the bulk to near the contact line to for a ring-like deposit (Figure 5b, also see Movie S3, ESI[3]).

Whereas, for the *heated* substrates, due to the presence of strong temperature gradient across the liquid-gas interface, Marangoni stresses establish the flow circulation from the contact line region to the droplet apex, with the formation of the axisymmetric Marangoni vortex, as shown schematically in Figure 5c [11, 38]. The particles get advected by this circulation and starts accumulating at droplet apex-point region. During the evaporation, an accumulation of the particles on the droplet top surface is seen in Figure 5d (see Movie S4, ESI[4]). In this context, the particles accumulation was also reported by Li et al. [28] and Parsa et al.[29].

---

[3]ESI, Movie S3 – A ring formation on glass at ambient temperature.

As pointed out by Li et al. [28] and shown by a schematic in Figure 5c, a stagnation-region exists near the pinned contact line due to three factors: radial outward flow near the contact line due to the largest evaporative flux on it, the Marangoni circulating flow, and curvature of the liquid-gas interface. The first factor leads to the deposition of few particles near the contact line, which results in the ring-formation. We also visualized such stagnation region in our experiments (see Movie S2, ESI[2]). At the end of the evaporation, the contact line depins and starts to recede asymmetrically with stick-slip motion [40] to form an inner deposit (see Movie S4, ESI4). Thus, on the heated substrates, near the pinned contact line, very few particles get deposited in ring and most of the particles get deposited as the inner deposit.

## 3.3 Effect of substrate wettability

The effect of substrate wettability is studied on the final deposition patterns considering almost *same* droplet volume (1.1±0.2 µL) on hydrophobic substrate (silicon wafer) as compared to the earlier hydrophilic substrate (glass). Note that the equilibrium contact angle is in range of 94-97° on the silicon wafer as compared to 34-38° on the glass substrate (Table 1). As seen in Figure 6a, during the evaporation at ambient temperature, initially (until $t \approx 96$ s) the contact line remains pinned and there is a decrease in the contact angle; thereafter, the contact line recedes at a constant contact angle (until $t \approx 480$ s). This is shown quantitatively in Figure 6c, with the initial-evaporation of the droplet in CCR mode (until $t = 110$ s) and later-evaporation in CCA mode. Thus, with increasing time, there is a *transition from the CCR to CCA mode of the evaporation* for the hydrophobic silicon substrate; in contrast to the CCR mode on the hydrophilic glass substrate during the complete evaporation (Figure 2c). The droplet surface isotherms and temperature profiles in Figure 6b and 6d, respectively, show smaller temperature gradients, similar to that on the glass substrate (Figure 2b and 2d). The Marangoni stresses are weak and the transition to the CCA mode of evaporation leads to an inner deposition of the particles (without ring formation) on the hydrophobic substrate, discussed later in this section.

---

[4]ESI, Movie S4 – Ring-like collection of particles during droplet evaporation on glass kept at 90°C temperature.

Figure 7 shows the droplet evaporation on heated, hydrophobic silicon substrate at $T_s = 60°C$ and $T_s = 90°C$. It is interesting to note that the *heating of the hydrophobic substrate* leads to the *suppression of the transition* from CCR to CCA mode or *delay in the depinning* of the contact line, and the occurrence of CCR mode of evaporation during most of the time of the evaporation, as shown in Figure 7(a, b). This delay in the depinning is explained as follows. In case of a pure liquid droplet, the depinning of the contact line occurs at the receding value of the contact angle, described by Young-Dupré equation [35].

In case of the droplet loaded with colloidal particles on the heated substrate, the particles are transported to the contact line by the radial outward flow and away from the contact line due to the Marangoni flow. Since the temperature gradient along the liquid-gas interface is larger for the heated substrate (Figure 7(c, d)) as compared to the non-heated (Figure 6d) case, the intensity of the Marangoni flow is significantly larger in the former. These two characteristics flows help in the formation of a stagnation region, similar to that seen on the glass in section 3.2. The trapped particles in this region offer resistance to the depinning and delay it. The phenomenon of the pinning the contact line by the particles themselves was named as "self-pinning" by Deegan [41].

The depinning is a function of substrate temperature since the intensity of convection inside the droplet and consequently the transport of the particles to the vicinity of the contact line increases with the substrate temperature. In order to quantify the delay in the depinning due to the substrate heating, experiments with pure liquid droplets were performed on the heated silicon. Figure 8 compares the depinning time, normalized with respect to the drying time, of the droplet containing colloidal particles ($c = 0.05\%$) with that of pure liquid droplet, for varying substrate temperature. The depinning is delayed for the former as compared to that in the latter at $T_s \geq 53°C$. Therefore, there exists a critical substrate temperature at which the delay of the depinning occurs, which is lower for the former (53°C) as compared to that in the latter (74°C). Note that the particles concentration also influences the delay, as discussed later in section 3.6.

The heating-induced delay in the depinning leads to the change in the shape of the deposit from *inner deposit without ring* to *inner deposit with ring*, shown in the top inset of Figure 9a. The figure shows a *slight* increase and decrease in the height and width of the ring, respectively, with increasing temperature of the hydrophobic silicon substrate, which is opposite to that observed on hydrophilic glass substrate (Figure 4a). Note that the variation in the height and width of the ring with the substrate temperature is *significant* at larger particles concentration (discussed later in section 3.4). This is explained due to larger contact angle and consequent larger Marangoni velocity ($V_{Ma}$, eq. 2) in case of the silicon as compared to that on the glass. $V_{Ma}$ for the former is plotted in Figure 9b and is one order of magnitude larger as compared to that on the glass (Figure 4b). In case of the silicon, the size of the stagnation region (Figure 10c) slightly decreases and increases in radial and axial direction, respectively, as compared to that in the glass (Figure 5c). Therefore, the particles self-assemble in several layers near the contact line on the silicon, resulting in a larger ring height and lesser ring width.

The droplet drying times are 540 s and 360 s on silicon (Figure 6c) and glass (Figure 2c), respectively, at $T_s = 27.5°C$ (ambient). On the other hand, these times are 80 s and 60 s for the former (Figure 7a) and latter (Figure 3a), respectively, at $T_s = 60°C$. Thus, the drying time in case of silicon is larger by 50% and 33% as compared to that for the glass at $T_s = 27.5°C$ and $T_s = 60°C$, respectively. These observations can be explained by the fact that the drying time scales with initial contact angle [42], which is larger in case of the silicon. However, the drying times are almost same for the silicon (Figure 7b) and glass (Figure 3b) at $T_s=90°C$, which is attributed to one order of magnitude larger Marangoni convection in the former as compared to that in the latter at $T_s=90°C$ (Figure 9b and Figure 4b).

The mechanisms of the formation of the deposit on the non-heated and heated hydrophobic silicon wafer are shown in Figure 10. On non-heated substrate, the contact line starts receding very early during the evaporation. Owing to the receding, the particles deposit in the inner region after the complete evaporation, as shown by a schematic in Figure 10a, and as observed in experimental results in Figure 10b (see also Movie S5, ESI[5]). On the other hand, in case of the heated substrate, the suppression of the transition (from CCR to CCA mode) occurs and the contact line almost remains pinned during most of the time of the evaporation. This is possibly due to the pinning of the contact line by particles 'trapped' in the stagnation region. This region forms due to the radially outward flow and Marangoni circulation, as shown using schematic in Figure 10c. The formation of the stagnation region near the contact line was also postulated by Li et al. [28]. Thus, the advection of the particles in the stagnation region by the radially outward flow and transport of the particles away from the stagnation region by Marangoni circulation form a ring and an inner deposit, respectively (see Movie S6, ESI[6]). Figure 10d shows the images of particles deposit formation for $c = 0.05\%$ solution on heated silicon kept at 90°C. The ring-like collection of particles on the droplet top surface form due to the Marangoni circulation, and finally at the last stage of evaporation the particles get deposited at the inner region with a thinner outer ring. Note that these deposits are similar to the deposits described in section 3.2 for the heated glass. The images at the top of Figure 9a (top side) show the final particle deposit patterns with the ring as well as the inner deposit, for the various temperatures of the silicon substrate.

---

[5]ESI, Movie S5 – Contact line depinning and particle deposit formation during droplet evaporation on hydrophobic silicon kept at ambient temperature.

### 3.4 Effect of particles concentration

In order to investigate the effect of the particles concentration on the final deposit shape, the ring profiles obtained at different substrate temperatures are compared in Figure 11 for two cases of particles concentration, $c = 0.1\%$ and 1%. The ring profiles are plotted for the glass and silicon substrates in top and bottom row, respectively. The microscopy images in the left column of each subfigure in Figure 11 show the final particles deposition pattern obtained at different substrate temperatures.

Figure 11a compares ring profiles obtained on the glass substrate at $c = 0.1\%$ at different substrate temperature ($T_s$). The thinning of the ring and formation of the inner deposit with increasing $T_s$ is due to stronger Marangoni convection in the evaporating droplet, as explained in section 3.2. Some particles deposit near the contact line due to stagnation region and majority of the particles move towards the droplet top surface due to the Marangoni circulation. Figure 11b plots the ring profiles at larger particles concentration, $c = 1\%$, at different $T_s$. Similar trends of the thinning of the ring with increasing $T_s$ are noted. As expected, at $c = 1\%$, more particles get accumulated in the ring as compared to $c = 0.1\%$ at corresponding substrate temperature, which results in larger width as well as height of the ring at all substrate temperatures. The quantification of the decrease in the ring mass with the substrate temperature as well as its scaling with the Marangoni velocity is presented in section 3.5.

---

[6]ESI, Movie S6 – Contact line pinning and particle deposit formation during droplet evaporation on hydrophobic silicon kept at 90°C temperature.

The ring profiles for the heated silicon substrate at $c = 0.1\%$ and $c = 1\%$ are plotted in Figure 11c and Figure 11d, respectively. The images in the left column in Figure 11 (c, d) show that final deposit as an inner deposit at $T_s = 27.5°C$ (ambient temperature) and a thin ring with an inner deposit at, $T_s = 60°C$ and $T_s = 90°C$. The dotted circle in the corresponding inset at $T_s = 27.5°C$ represents the initial wetted diameter of the droplet. In Figure 11(c, d), the height of the ring increases with increasing substrate temperature ($T_s$) due to increasing Marangoni velocity ($V_{Ma}$), as explained in section 3.3 (Figure 9b). At larger particles concentration, more particles get accumulated at the droplet top surface and consequently start depositing at the droplet inner region. This can be qualitatively confirmed from the images of the dried deposit in the left column of Figure 11(c, d).

The comparisons of the ring profiles at $c = 0.1\%$ as well as $c = 1\%$ between heated hydrophobic silicon and heated hydrophilic glass show substantial increase in the height of the ring for the former as compared to that in the latter at the corresponding substrate temperature. This is probably due to much larger contact angle as well as Marangoni velocity $V_{Ma}$ for the evaporation on the silicon substrate. We qualitatively observe that the area of the inner deposit at the corresponding substrate temperature increases with increase in the particles concentration on both substrates, as shown by the images of deposition pattern in Figure 11.

## 3.5 Comparison of measurements with first-order modeling

In this section, we compare experimental results with the predictions by first-order models, proposed below in separate subsections for the temperature measurements at the droplet apex-point and distribution of the mass of the particles in the ring.

### 3.5.1 Temperature at the droplet apex

The measured temperature at the droplet apex (shown as a red dot in Figure 12a) is compared with the values obtained in the modeling. A quasi-steady, one-dimensional, heat conduction model is coupled with a quasi-steady, diffusion-limited evaporation to obtain the temperature at the apex (Figure 12a). The validity of the quasi-steady evaporation is justified [7] if $t_{heat}/t_f < 1$, where $t_{heat}$ and $t_f$ are heat equilibration time of the droplet and droplet drying time, respectively. The former is given by $t_{heat} \sim h_L^2/\alpha$, where $h_L$ and $\alpha$ are the droplet height and thermal diffusivity of the droplet, respectively and the latter is expressed as, $t_f \sim 0.2(\rho_L/\rho_{vap})(r_{w,i}h_L/D_{vap})$, where $\rho_L$, $\rho_{vap}$, $r_{w,i}$ and $D_{vap}$ are liquid density, vapor density, initial wetted radius of the droplet and liquid vapor diffusion coefficient, respectively. The maximum value of $t_{heat}/t_f$ for the cases considered in the present work is around 0.05 for the heated silicon at $T_s = 90°C$, justifying the quasi-steady assumption. A one-dimensional control volume considered in the droplet and substrate is shown in Figure 12a. Convection inside as well as outside the droplet is neglected and a perfect thermal contact between the droplet and the substrate is assumed. The energy conservation equation along droplet and substrate thickness in the control volume with these assumptions simplifies to,

$$\frac{d^2T}{dz^2} = 0 \qquad (3)$$

The boundary condition at the bottom surface of the substrate is,

$$T(z=0) = T_s \qquad (4)$$

The boundary condition at the top boundary of the control volume is jump energy boundary condition and is given by,

$$J_{apex} L = -k_L \frac{dT}{dz} \quad (5)$$

where, $J_{apex}$ is the evaporative mass flux at the apex of the droplet [kg/m²-s], $L$ is the latent heat of the evaporation of the liquid [J/kg], and $k_L$ is the liquid thermal conductivity [W/m-K]. The evaporative mass flux $J$ at the liquid-gas interface for contact angle $\theta \leq 90°$ is expressed as [39],

$$J(r,T) = J_0(\theta,T) \left[ 1 - \left( \frac{r}{r_{w,i}} \right)^2 \right]^{-\lambda(\theta)} \quad (6)$$

where $r$, $\theta$, $r_{w,i}$ are the radial coordinate, contact angle and initial wetted radius respectively. The expression for $J_0(\theta)$ is given by [39]:

$$J_0(\theta,T) = \frac{D_{vap}(T)(c_{sat}(T) - Hc_\infty)}{R}(0.27\theta^2 + 1.30)(0.6381 - 0.2239(\theta - \pi/4)^2) \quad (7)$$

The variation of diffusion coefficient ($D_{vap}$) and liquid vapour saturated concentration ($c_{sat}$) with temperature are given by following expressions [11, 43, 44],

$$D_{vap}(T) = (2.5e-4) e^{-\frac{684.15}{(T+273.15)}} \quad (8)$$

$$c_{sat}(T) = [9.99 \times 10^{-7} T^3 - 6.94 \times 10^{-5} T^2 + 3.20 \times 10^{-3} T - 2.87 \times 10^{-2}] \quad (9)$$

The value of $J_{apex}$ is defined at $r = 0$, as follows,

$$J_{apex} = J_0(\theta, T_{apex}) \quad (10)$$

The solution of eq. 3 using boundary conditions (eqs. 4 and 5) is given by [22],

$$T_{apex} = T_s - J_{apex} L \left( \frac{h_S}{k_S} + \frac{h_L}{k_L} \right) \quad (11)$$

where, $h_S$ is substrate thickness, $k_S$ is thermal conductivity of the substrate, $h_L$ is droplet height, and $k_L$ is thermal conductivity of the droplet. Since $J_{apex}$ is a function of $T_{apex}$ in eq. 11, $T_{apex}$ is calculated by an iterative-solution of eqs. 6-11.

The measurements of the heated glass and silicon at $T_s = 60°C$ in Figure 12b show a large increase in $T_{apex}$ within first few seconds and then remains constant for quite some time (shown by a dotted horizontal line). This plateau value of experimentally obtained $T_{apex}$ is considered for comparisons with the quasi-steady model. A good agreement is shown in Figure 12c, between the measurements and the values obtained by the model and errors are within 1-3% and 4-8% for the glass and silicon, respectively. The larger errors for the silicon may be attributed to larger intensity of the Marangoni convection which is neglected in the model.

### 3.5.2 Particles deposition in the ring

For the mass of the particles deposited in the ring, a scaling analysis is presented considering the characteristics velocities in the evaporating droplet. In this regard, percentage of the mass of the particles present in the ring with respect to the total mass of the particles is defined as,

$$\lambda(\%) = \frac{\text{Mass of particles present in the ring }(M_{ring})}{\text{Total mass of the particles }(M_{total})} \times 100 \qquad (12)$$

where $M_{ring}$ is estimated by fitting a second-order curve to the measured ring profile and integrating it with respect to the radial distance. A typical fitting is shown in the supplementary information. Thus, the expression of $M_{ring}$ is given as follows,

$$M_{ring} = \int_{r_{w,i}-\Delta r}^{r_{w,i}} f(r) 2\pi r \, dr \qquad (13)$$

where $f(r)$, $r_{w,i}$ and $\Delta r$ are the fitted curve, initial wetted radius and ring thickness, respectively.

The present scaling analysis is based on the two characteristic velocities namely, radial outward velocity $V_{rad}$ (due to the maximum evaporation at the contact line) and the Marangoni velocity $V_{Ma}$ (generated due to the temperature difference across the liquid-gas interface $\Delta T$). We postulate that the $V_{rad}$ transports the particles near the contact line while $V_{Ma}$ takes them away from the contact line, shown schematically in Figure 13a. Thus, net mass transported near the contact line in the ring scales as, $V_{rad}$ - $V_{Ma}$; where $V_{Ma}$ is estimated using eq 2 (where $\Delta T$ is based on the plateau values; section 3.5.1) and $V_{rad}$ scales as [12],

$$V_{rad} \sim \frac{J_{max}}{\rho_L} \quad (14)$$

where $J_{max}$ is the maximum evaporative flux near the contact line (eq. 6) and $\rho_L$ is the droplet density.

Figure 13b shows variation of $\lambda$ with substrate temperature for the hydrophilic glass at three cases of particles concentration ($c$ = 0.05%, 0.1% and 1%, v/v), together with $V_{rad}$ - $V_{Ma}$ at different substrate temperatures. The values of $V_{rad}$ - $V_{Ma}$ are independent of the particle concentration (eqs. 2 and 14). The percentage of the particle deposition in the ring, $\lambda$, shows a non-monotonic decrease with substrate temperature at different cases of the particles concentration. $V_{rad}$ - $V_{Ma}$ is negative for all cases of substrate temperature since $V_{Ma} > V_{rad}$ and its absolute magnitude increases with increasing substrate temperature; along with a decrease in the percentage of the particle deposition in the ring. Therefore, it verifies the hypothesis that the larger Marangoni velocity (as compared to the radial velocity) for the heated substrates takes away more particles from contact line region to droplet inner region (Figure 13a). Thus, $\lambda$ scales with $V_{rad}$ - $V_{Ma}$ and inversely scales with the substrate temperature. Figure 13b also plots similar trend of the variation for $\lambda$ values reported by Li et al. [28], on a heated ($T_s$ = 30 to 80°C) hydrophilic glass substrate at $c$ = 0.25% (v/v) for 2.5 µL water droplet, consistent with our model predictions.

Similar figure for the heated hydrophobic silicon is shown in Figure 13c, with smaller $\lambda$ values (as compared to Figure 13b) at $T_s = 60°C$ and $T_s = 90°C$ - due to the formation of thin ring with an inner deposit. Note that $\lambda$ is not defined at $T_s = 27.5°C$ at all particles concentrations since the ring does not form in these conditions. With temperature increasing from 60°C to 90°C, the figure shows that $\lambda$ remains almost constant; however, there is a large decrease in $V_{rad}$ - $V_{Ma}$. Thus, the present model predictions do not match with the measurements in case of the heated silicon and this discrepancy may be explained as follows. The mass of the particles in the ring (~10% - 20%) is on the order of experimental uncertainty and therefore, a higher-order model is required to predict the measurements in case of the heated silicon.

## 3.6   Regime map

We propose a regime map for predicting the shape of the deposit obtained after the evaporation of a droplet containing colloidal particles. The map in Figure 14 demarcates three regimes for the deposits, namely, ring, a thin ring with inner deposit and inner deposit, obtained as function of substrate temperature, contact angle and particles concentration. The insets show the images of the deposit obtained by optical microscopy measurements. The measurements reported by Nguyen et al. [26] and Li et al. [28] are also plotted in this map.

The first regime represents *ring* which forms at low substrate temperature and contact angle. A typical image of the deposit is shown in inset (bottom left) of Figure 14. This deposit forms by a dominating radially outward flow due to the maximum evaporation near the pinned contact line. The formation of the ring-like deposit has been extensively reported in the literature.

The second regime represents *thin ring with inner deposit* which forms at larger substrate temperature for a wide range of the contact angle. A typical image of the deposit is shown in the insets (right) of Figure 14. It is attributed to the presence of strong Marangoni convection in the evaporating droplet, which transports the particles in the inner region of the particles. Since the intensity of the Marangoni convection scales with the substrate temperature, the ring becomes thinner at larger substrate temperature. This deposit is also reported at larger particles concentration, as plotted in Figure 14. Measurements of Li et al.[26] for the heated hydrophilic glass substrate also plotted within this regime in Figure 14.

The third regime represents *inner deposit* which forms at larger contact angle and lower substrate temperature. A typical image of the deposit is shown in inset (top left) of Figure 14. In these conditions, contact line starts receding very early and the evaporation occurs in constant contact angle (CCA) mode. Thus, all particles are deposited in an inner deposit, which is much smaller in size as compared to the initial wetted diameter. The substrate heating aids in pinning of the contact line due to self-pinning (as explained in section 3.3), which results in a thin ring with an inner deposit at larger substrate temperature. A larger particle concentration also aids in the pinning of the contact line and thus, there exists a critical substrate temperature and critical particles concentration for the transition from third (inner deposit) to the second (thin ring with inner deposit) regime. Therefore, the boundary between the third and second regime depends on the substrate temperature as well as particle concentration, as plotted by straight broken lines in Figure 14. Measurements of Nguyen et al.[26] for the substrate at ambient temperature are also plotted within third regime in Figure 14.

## 4  Conclusions

An experimental investigation of the evaporation of a sessile water droplet containing colloidal particles is performed. The effects of the substrate temperature, substrate wettability and particle concentrations on the evaporation dynamics and shapes of the particles deposit are quantified. Time-varying droplet shapes are recorded from side using high-speed visualization, and unsteady temperature of the liquid-gas interface is measured from top using infrared thermography. The motion of the particles inside the evaporating droplet is qualitatively visualized using an optical microscope and the profile of the ring-like deposit is measured using an optical profilometer. The hydrophilic glass and hydrophobic silicon substrates are maintained at 27.5°C, 60°C and 90°C and three cases of the particles concentration (0.05%, 0.1% and 1%, v/v) are considered.

On the hydrophilic glass, the contact line is pinned throughout the evaporation in constant contact radius (CCR) mode on the non-heated as well as heated substrate. In the *non-heated* case, the particles are advected to the contact line by the radially outward flow, developed due to the largest evaporative mass flux near the contact line and the final deposit is a *ring*. However, in the heated case, the Marangoni convection advects most of the particles to the droplet inner region and a stagnation region develops near the contact line. Consequently, final deposit is a *thin ring with inner deposit* on the *heated* glass. The width as well as height of the ring decreases with increase in the substrate temperature due to larger Marangoni velocity.

On the hydrophobic silicon, the droplet evaporates mostly in constant contact angle (CCA) in the *non-heated* case and an *inner deposit* forms in this case. In the heated case, the droplet evaporates mostly in the CCR mode and the heating of the substrate leads to the suppression of the transition from CCR to CCA mode or delay in the depinning of the contact line. The delay is explained by the self-pinning of the particles in the stagnation region; formed due to the radially outward flow, Marangoni circulation and curvature of the liquid-gas interface. There exists a critical substrate temperature as well as particles concentration at which the delay of the depinning of the contact line occurs. As a result, a *thin ring with inner deposit* forms on the *heated* silicon. The ring height in the heated case increases with increase in the substrate temperature, due to larger contact angle and larger Marangoni velocity.

Finally, we propose a regime map for predicting the deposit shape as function of substrate temperature, substrate wettability and particle concentration. The map demarcates the above-mentioned deposits namely, *ring*, *thin ring with inner deposit* and *inner deposit*. We formulate first-order models to verify the temperature measurements and particles deposition patterns. The present systematic measurements may serve as a good dataset to test and validate high-fidelity models in future. Overall, the present study helps to understand the thermal Marangoni effect inside the droplet coupled with the substrate wettability and the particles concentration of the colloidal solutions. The findings may aid in controlling the shape of the particles deposit in applications such as inkjet printing and self-assembly in bioassays.

## 5  Acknowledgements

R.B. gratefully acknowledges financial support from Department of Science and Technology (DST), New Delhi through fast track scheme for young scientists and an internal grant from Industrial Research and Consultancy Centre (IRCC), IIT Bombay. N.D.P. was supported by a Ph.D. fellowship awarded by IRCC. We thank two anonymous reviewers for the useful comments.

# 7 Figures

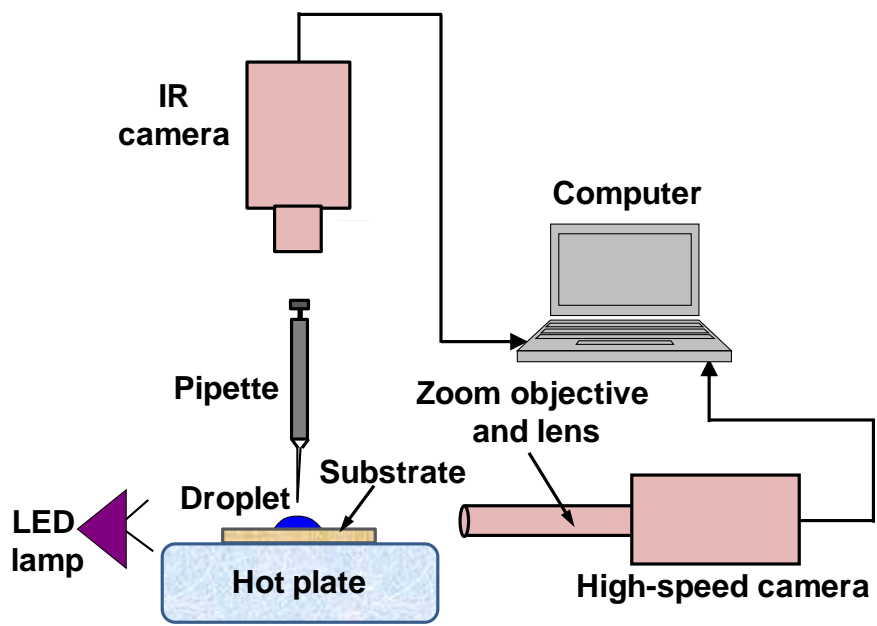

Figure 1: Schematic of the experimental setup.

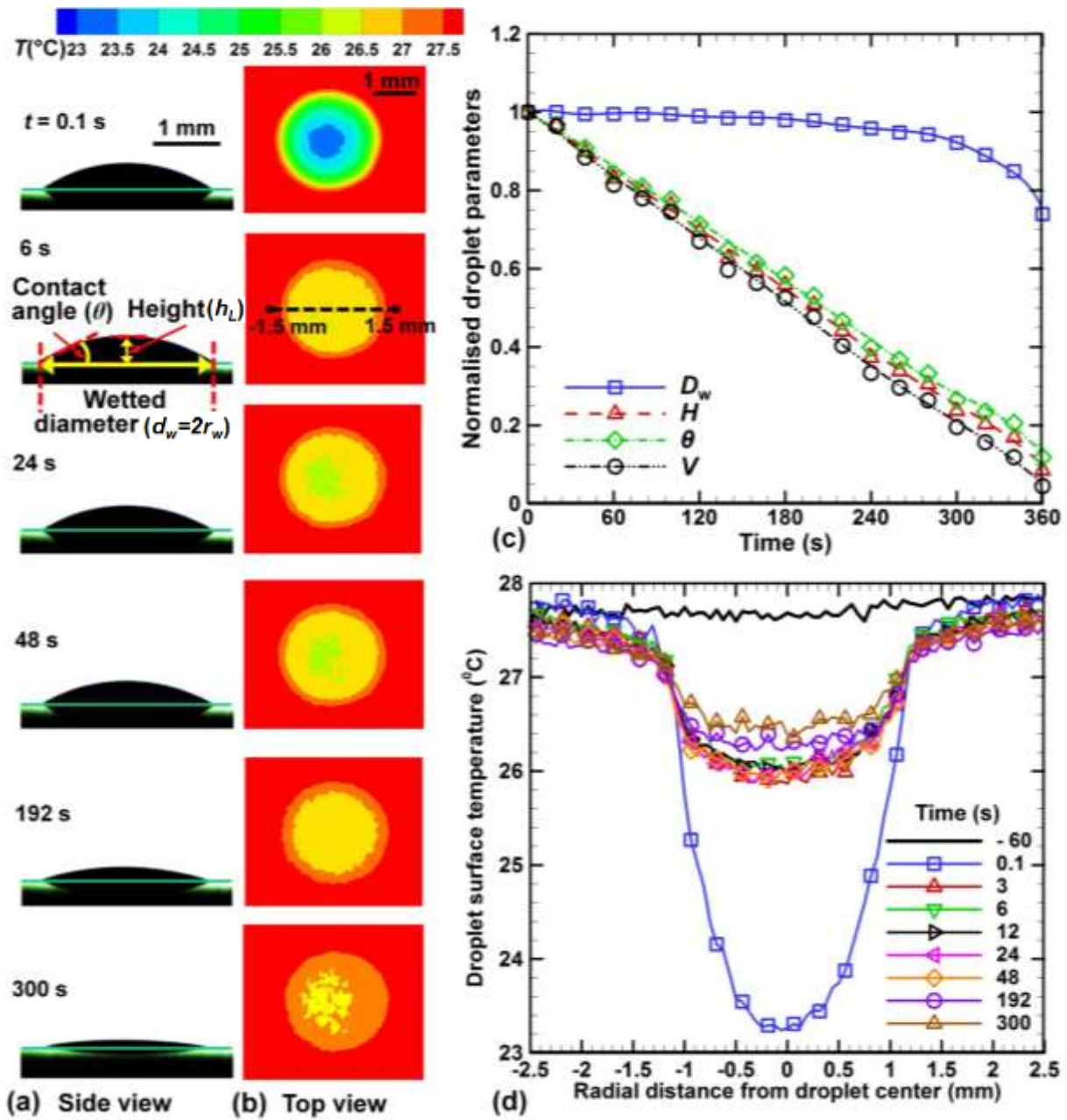

Figure 2: Evaporation of a 1.0 µL sessile droplet (containing 0.05% v/v colloidal particles) on non-heated glass substrate: (a) High-speed visualization from side of the droplet for different evaporation times. (b) Infrared thermography from top of the droplet showing the instantaneous isotherms on liquid-gas interface and substrate surface. (c) Temporal variation of wetted diameter, height, contact angle and volume. Temporal resolution for the data is 0.1 s. (d) Droplet surface temperature distribution profiles at different time instances with spatial resolution of 0.0625 mm.

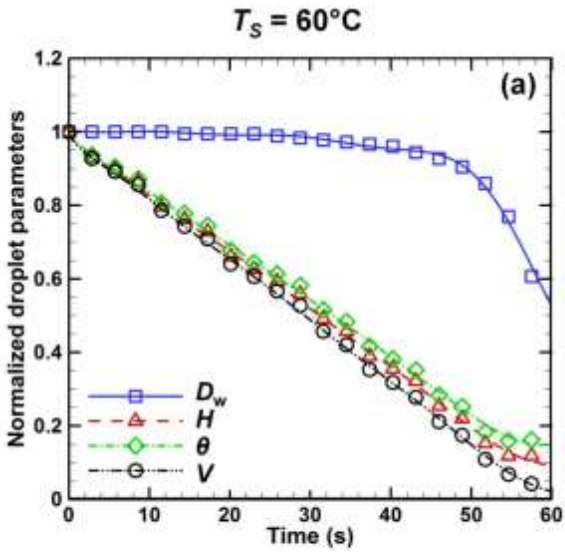
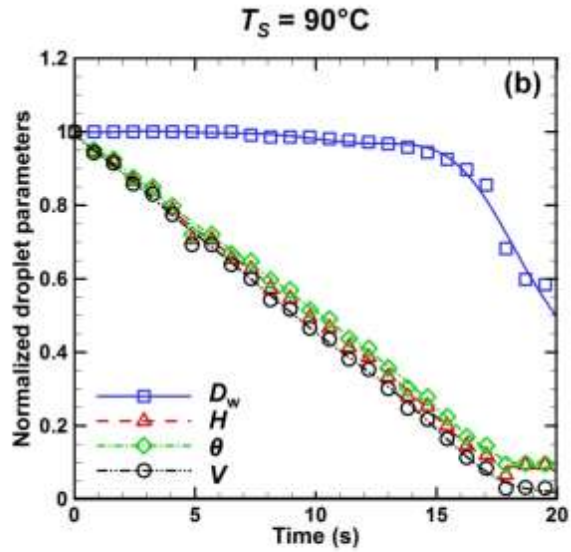
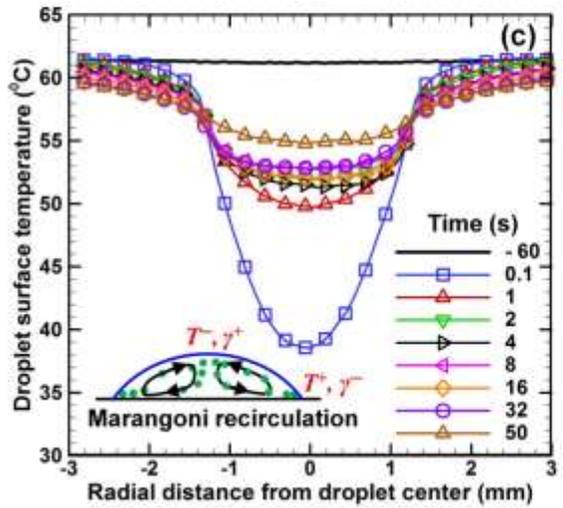
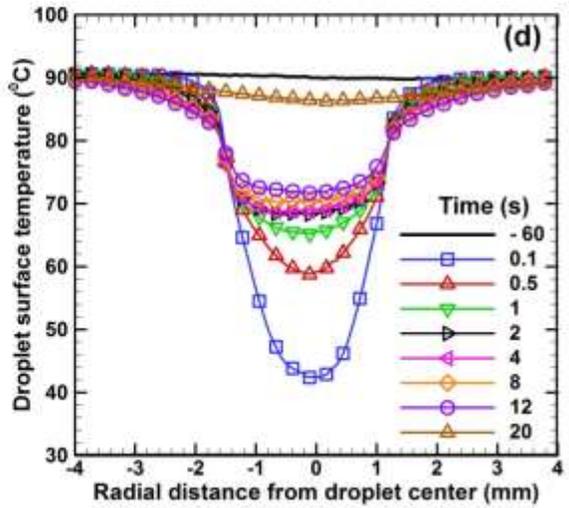

Figure 3: Evaporation dynamics and heat transfer for a 1.1 μL sessile droplet (containing 0.05% v/v colloidal particles), on the heated glass substrate maintained at (a, c) 60°C and (b, d) 90°C. (a, b) Temporal variation of wetted diameter, height, contact angle and volume. The temporal resolution for the data is 0.01 s. (c, d) Droplet surface temperature distribution profiles at different time instances with spatial resolution of 0.0625 mm.

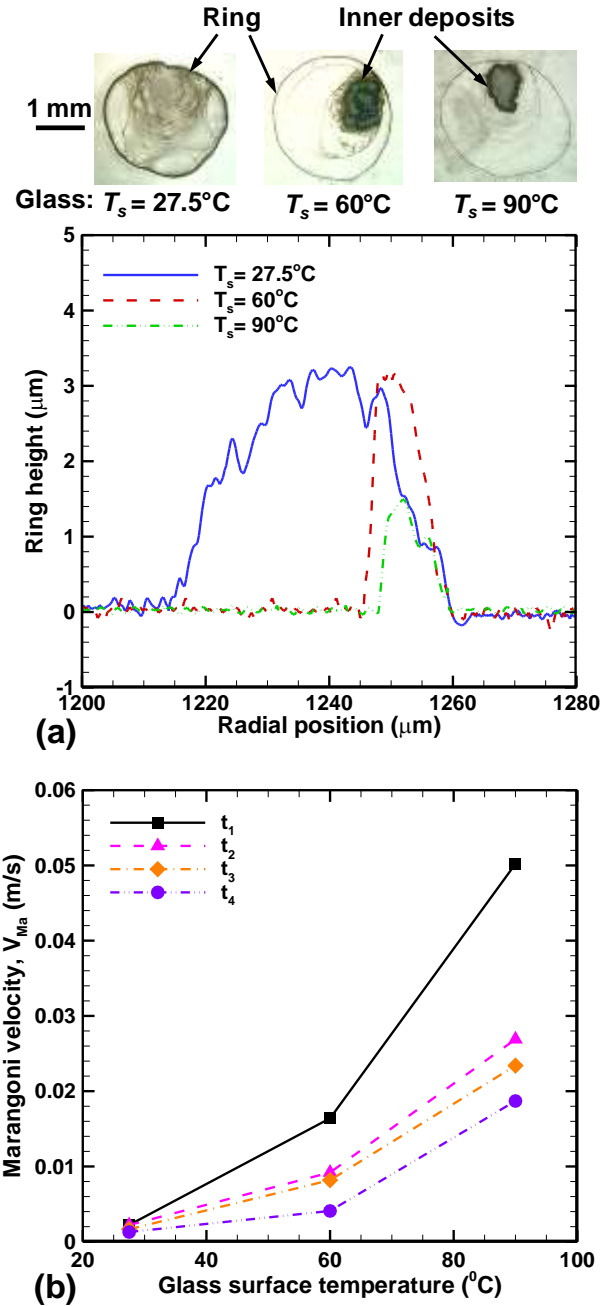

Figure 4: (a) Comparison of the measured ring profiles for the droplet ($c$ = 0.05%) evaporation on glass, maintained at different temperatures. The insets at the top show the top view of the dried deposit obtained. (b) Variation of Marangoni velocity $V_{Ma}$, with substrate temperature, at different time instances given in Table 2. For the computation of $V_{Ma}$ (eq. 2), the temperature-differences, $\Delta T$, between the droplet contact line and apex point are calculated from the plots shown in Figure 2d and Figure 3(c, d).

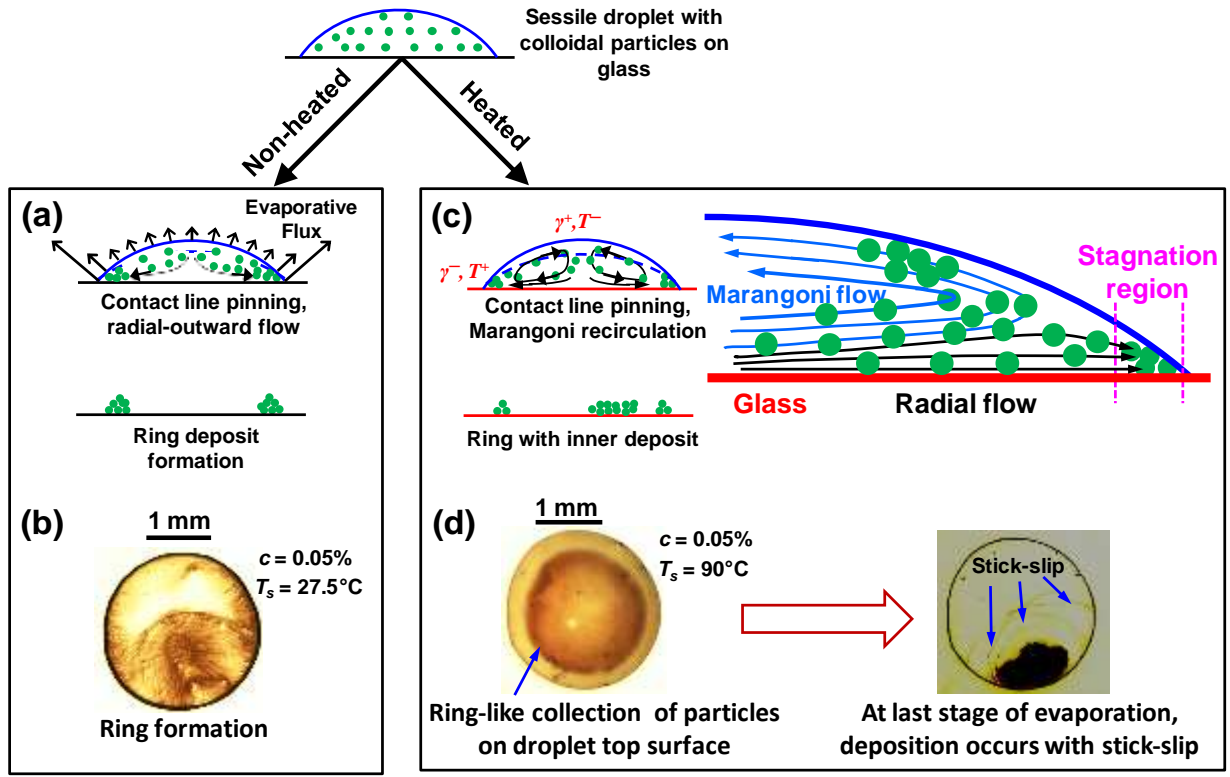

Figure 5: (a, c) Schematic representation of the evaporation-induced fluid flow and final particles deposit patterns on non-heated (a) and heated (c) hydrophilic substrate. (b, d) Experimental results of the deposit patterns on non-heated (b) and heated (d) hydrophilic glass substrate.

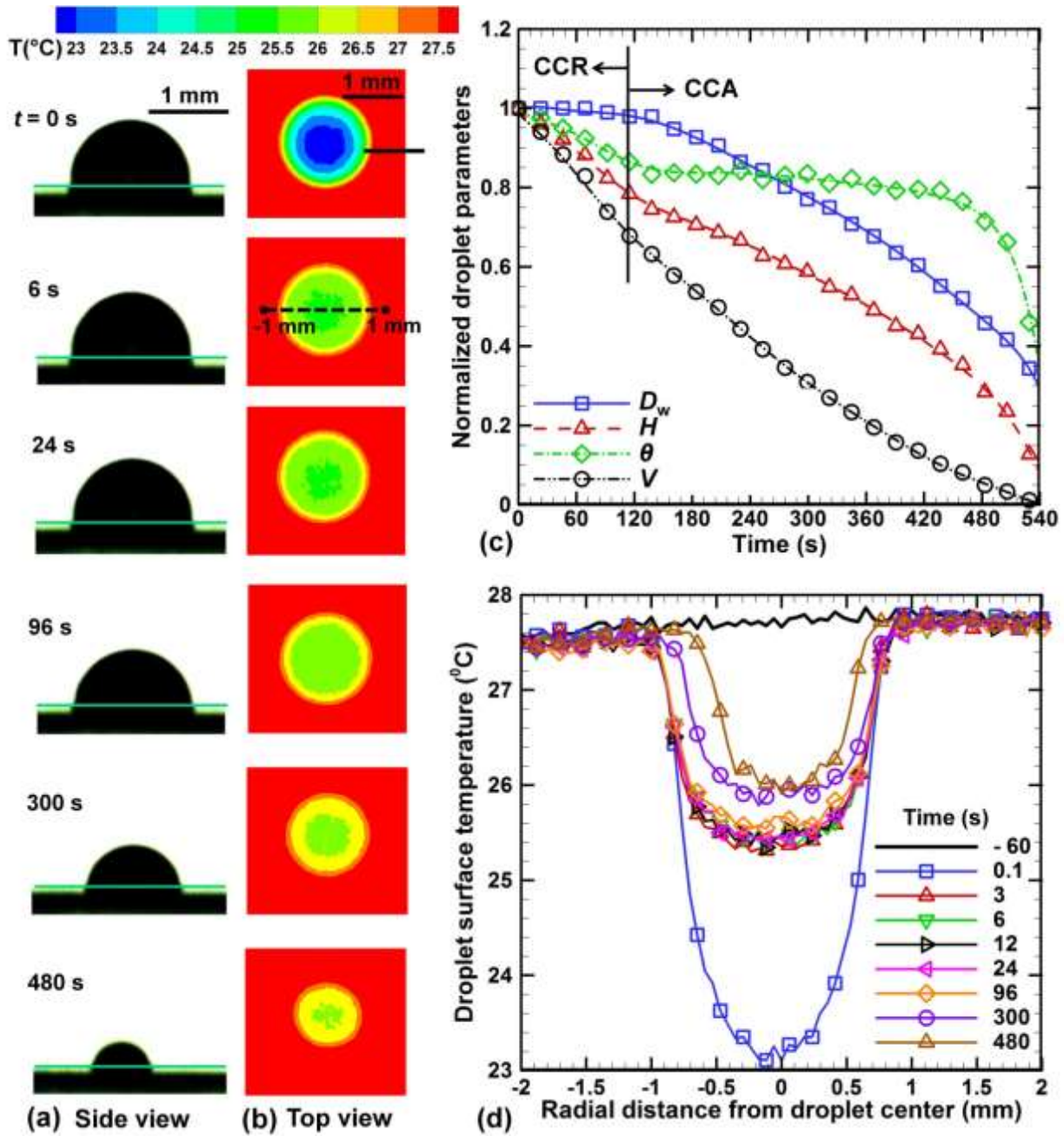

Figure 6: Evaporation of a 1.1 µL sessile droplet (containing 0.05% v/v colloidal particles) on non-heated silicon substrate: (a) High-speed visualization from side of the droplet for different evaporation times. (b) Infrared thermography from top of the droplet showing the instantaneous isotherms on liquid-gas interface and substrate surface. (c) Temporal variation of wetted diameter, height, contact angle and volume. Temporal resolution for the data is 0.1 s. (d) Droplet surface temperature distribution profiles at different time instances with spatial resolution of 0.0625 mm.

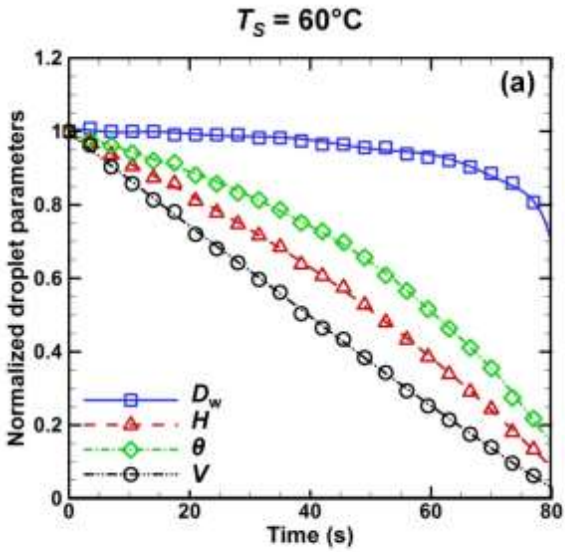
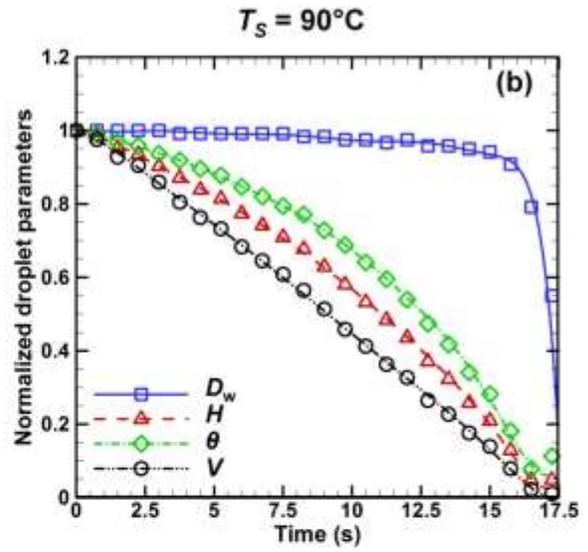
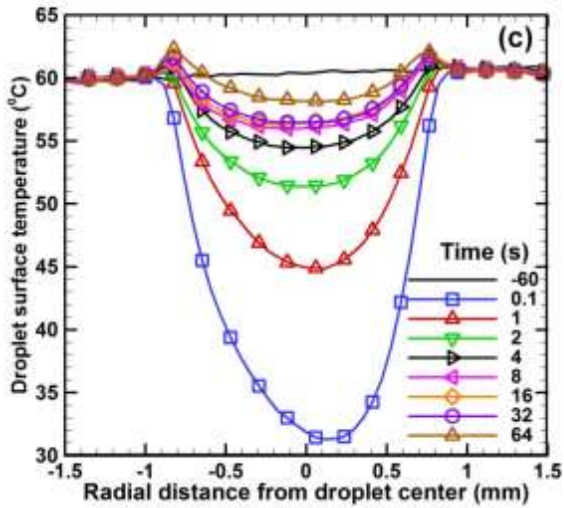
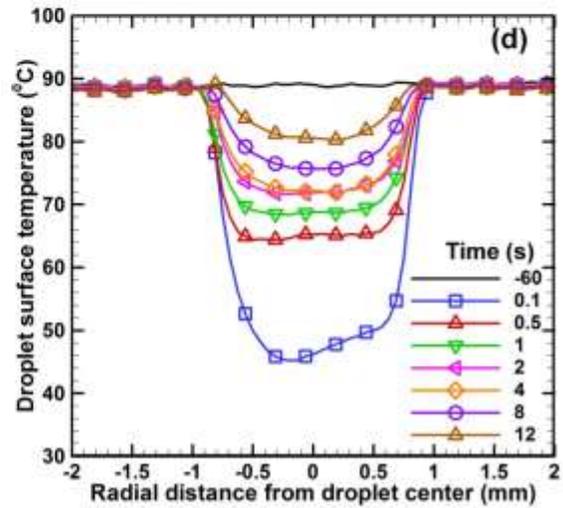

Figure 7: Evaporation dynamics and heat transfer for a 1.1 µL sessile droplet (containing 0.05% v/v colloidal particles), on the heated silicon substrate; maintained at (a, c) 60°C and (b, d) 90°C. (a, b) Temporal variation of wetted diameter, height, contact angle and volume. The temporal resolution for the data is 0.01 s. (c, d) Droplet surface temperature distribution profiles at different time instances with spatial resolution of 0.0625 mm.

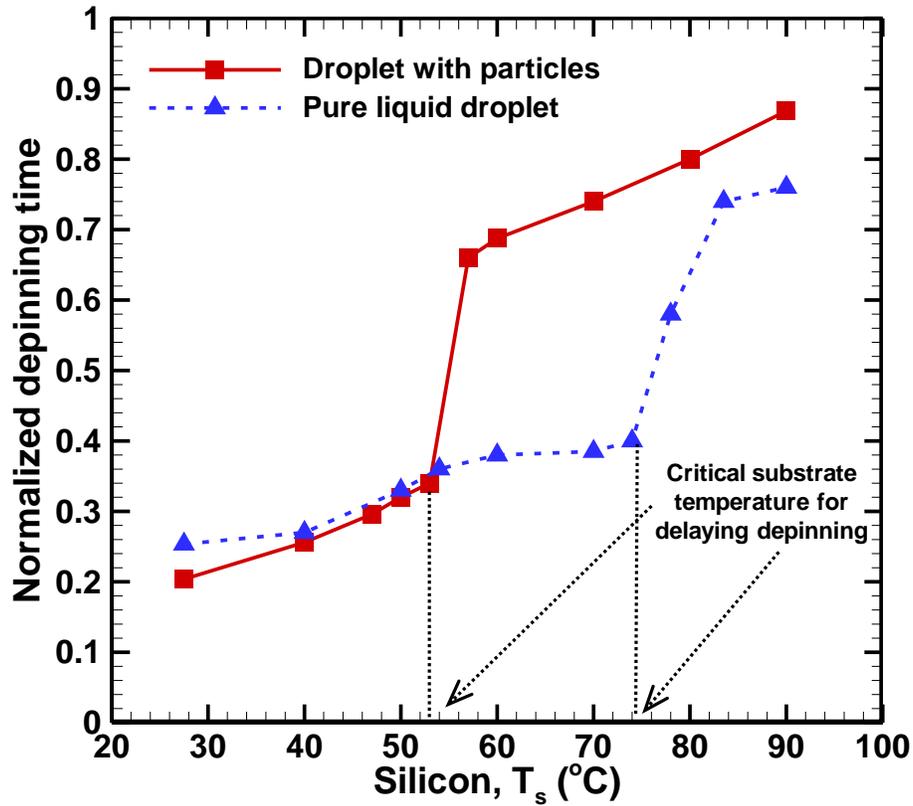

Figure 8: Normalized depinning time of the contact line during the evaporation of the pure liquid droplets and droplets containing colloidal particles (0.05% v/v) on heated hydrophobic silicon. The critical substrate temperature for the delayed depinning is larger for the former as compared to that for the latter.

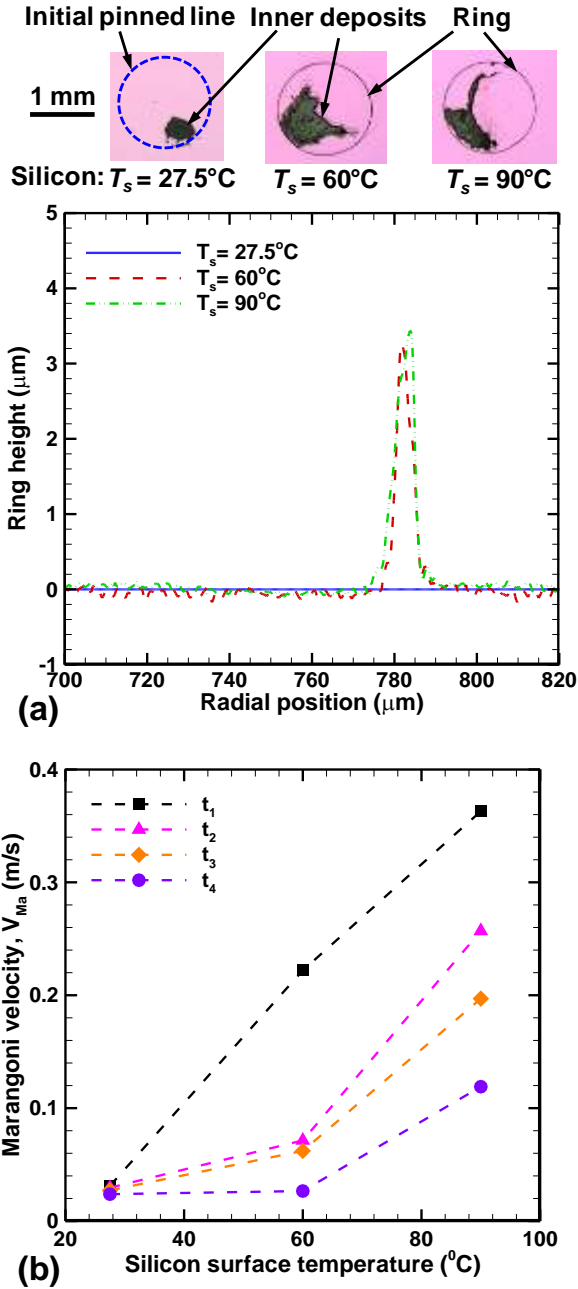

Figure 9: (a) Comparison of the measured ring profiles for the droplet ($c$ = 0.05%) evaporation on silicon, maintained at different temperatures. The insets at the top show the top view of the dried deposit obtained. (b) Variation of Marangoni velocities with substrate temperature at different time instances given in Table 2. For the computation of $V_{Ma}$ (eq. 2), the temperature-differences, $\Delta T$, between the droplet contact line and apex point are calculated from the plots shown in Figure 6d and Figure 7(c, d).

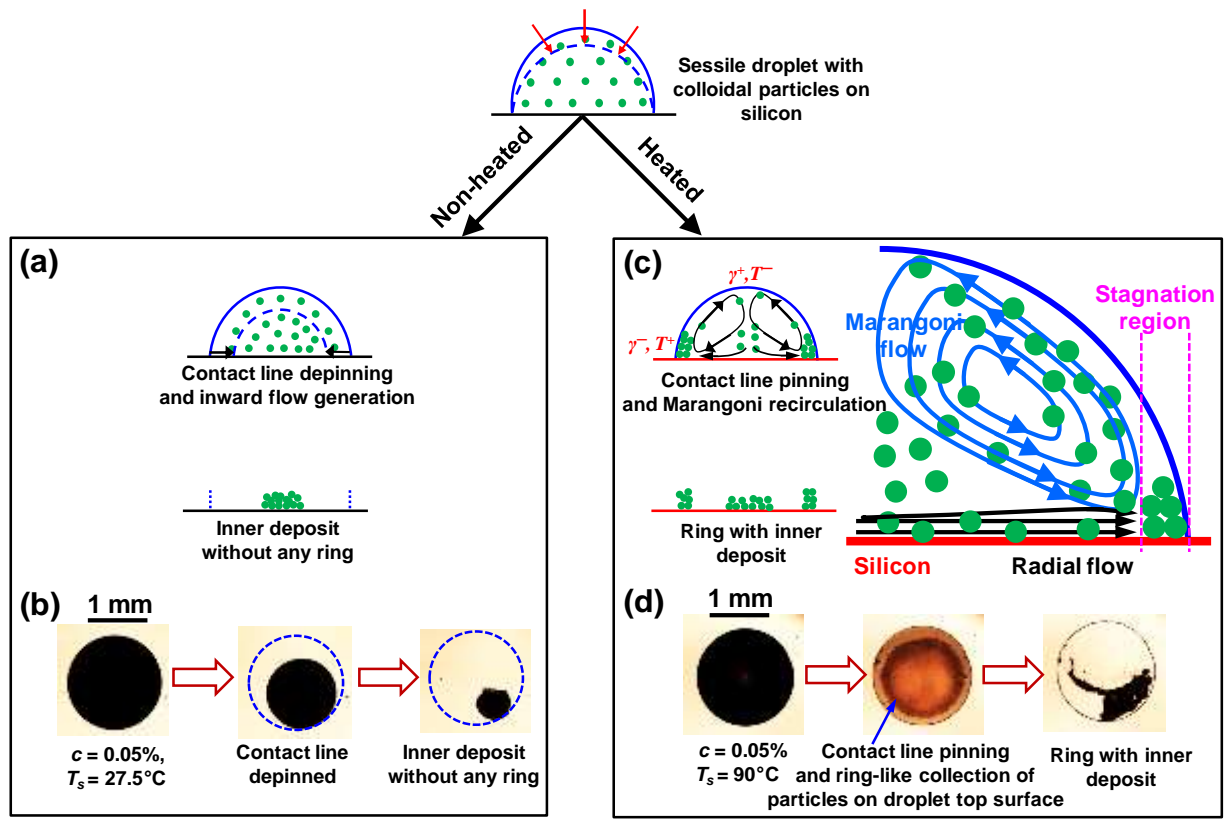

Figure 10: (a, c) Schematic representation of the evaporation-induced fluid flow and final particles deposit patterns on non-heated (a) and heated (c) hydrophobic substrate. (b, d) Experimental results of the deposit patterns on non-heated (b) and heated (d) hydrophobic silicon substrate.

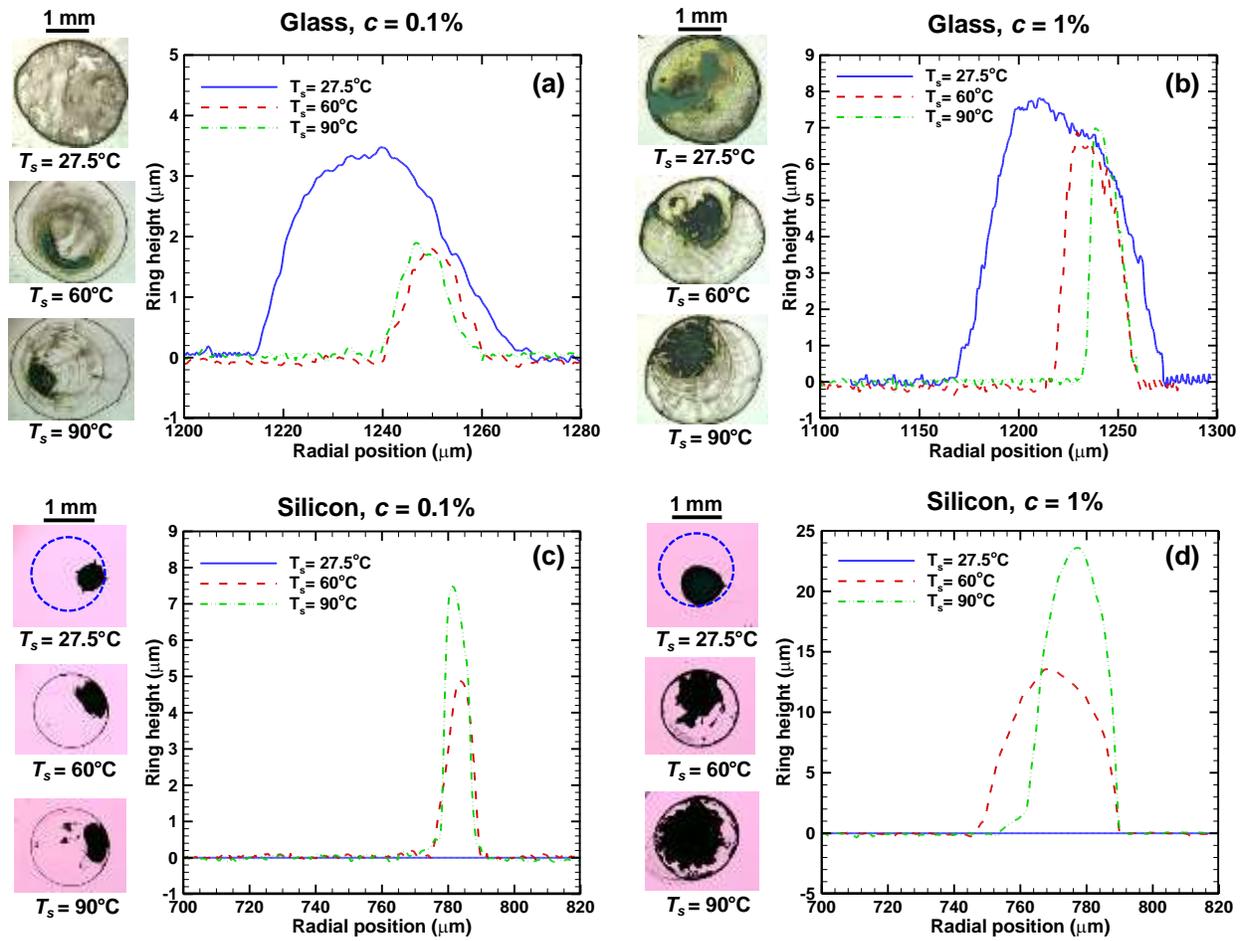

Figure 11: Comparison of the measured ring profiles obtained after evaporation of 1.1±0.2 µL droplets on glass (a, b) and silicon (c, d) substrates maintained at different temperatures. The particles volume concentrations are 0.1% (a, c) and 1.0% (b, d).

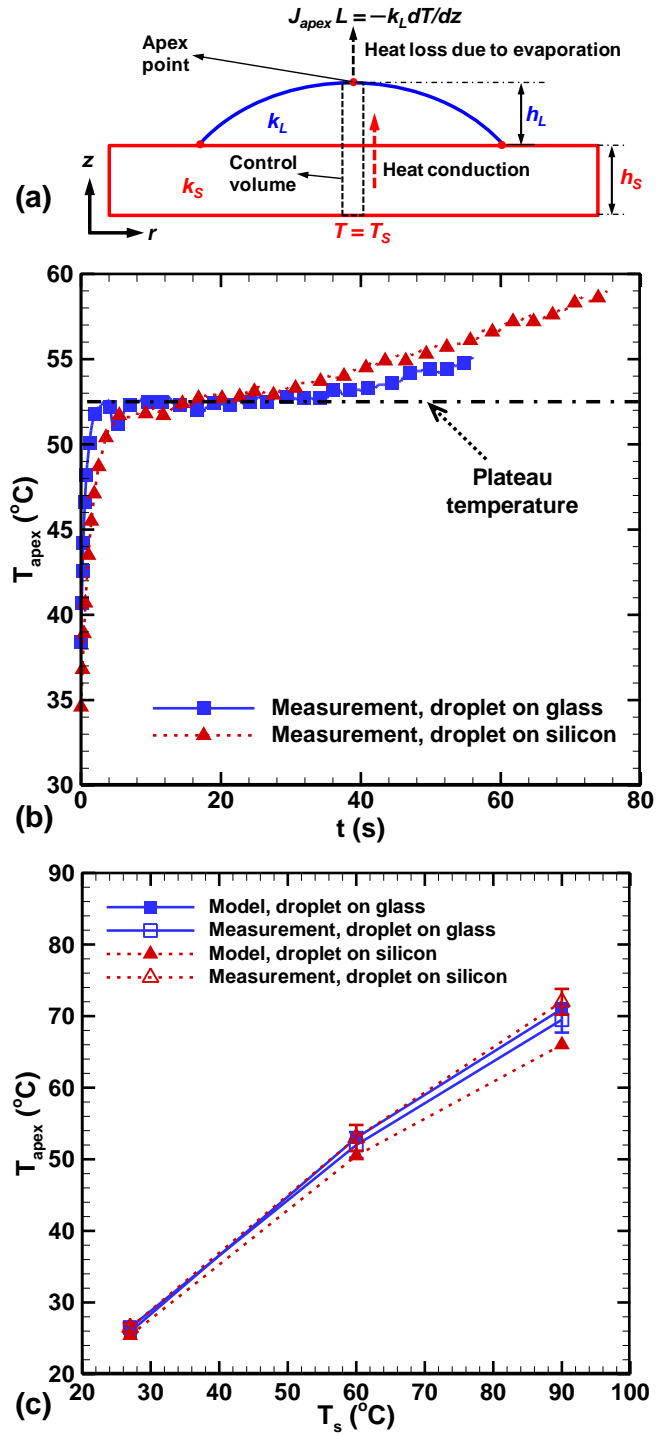

Figure 12: (a) Schematic of a one-dimensional heat transfer model for an evaporating sessile droplet on a heated substrate. (b) Measured plateau temperature using IR camera at the droplet apex during evaporation on glass and silicon substrates for $T_s = 60°C$. (c) Comparisons of droplet apex temperature ($T_{apex}$) obtained by the model (eq. 11) with the measurements at different substrate temperatures ($T_s$).

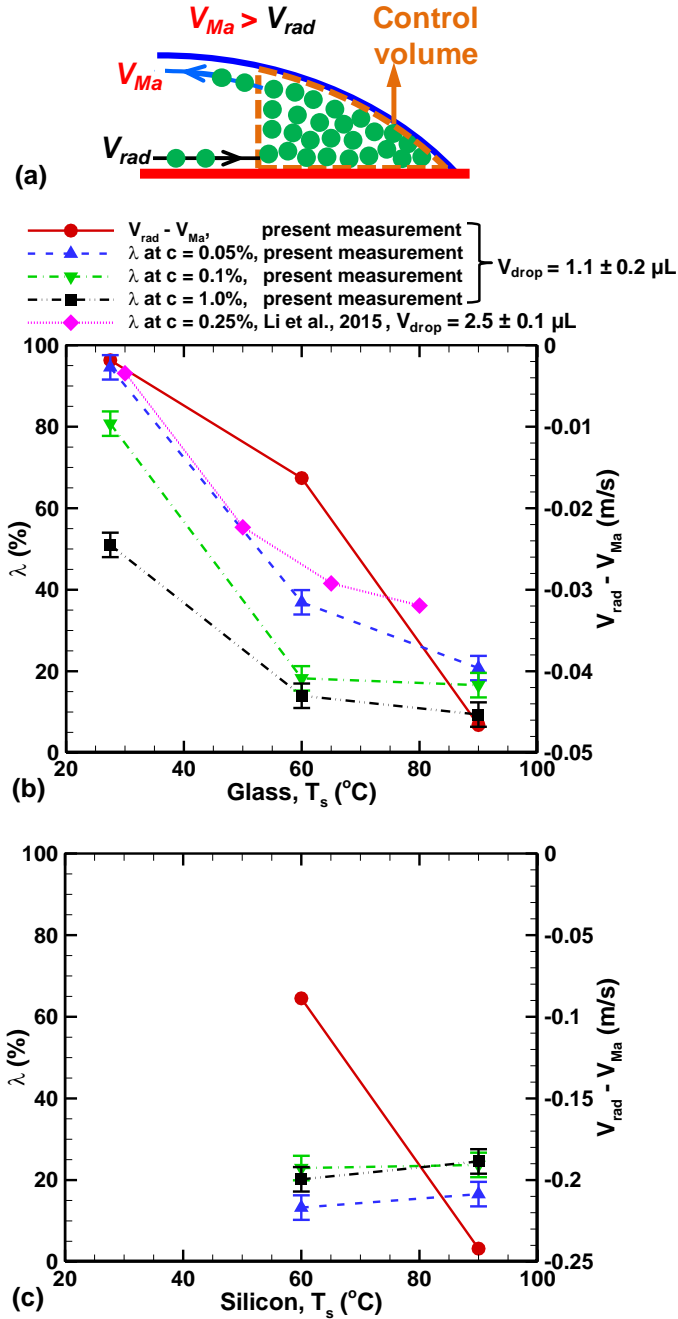

Figure 13: (a) Control volume approach for estimating the mass of particles deposition near the contact line. (b, c) Comparison of percentage of the mass of particles deposited in the ring, $\lambda$, for different particles concentration in the present measurements and measurements of Li et al. [28] for 2.5 μL droplet. The characteristic velocity, $V_{rad} - V_{Ma}$, is also plotted for the droplet on glass (b) and hydrophobic silicon (c). Note that $\lambda$ for the silicon at 27.5°C is not defined because the ring does not form at this temperature in all cases of the particles concentration.

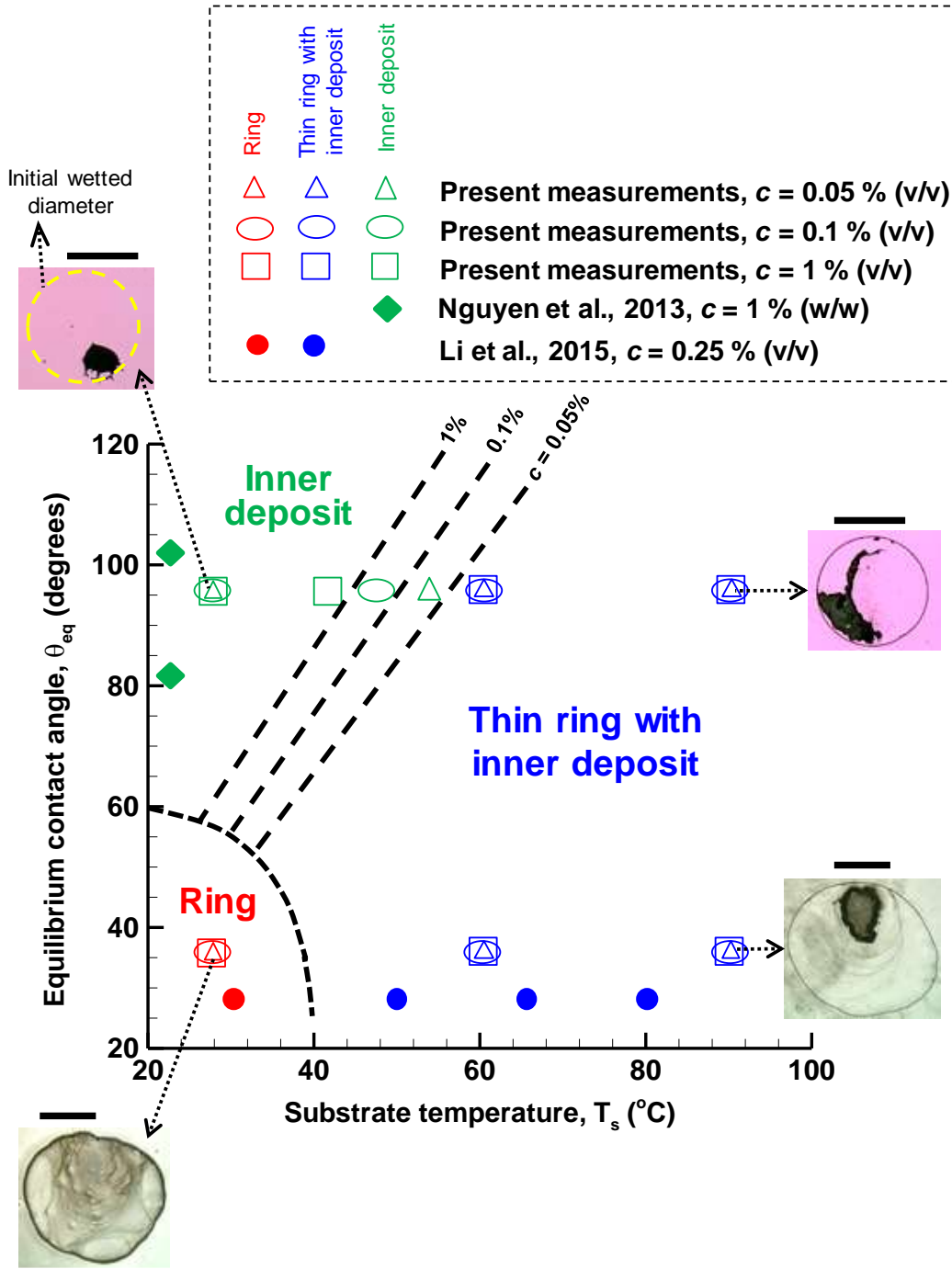

Figure 14: Regime map for predicting the deposit shape as function of the substrate temperature, substrate wettability and particles concentration. The dashed lines are plotted to demarcate the three regimes. Non-filled and filled symbols represent present and published measurements, respectively. Insets show the deposit image obtained by optical microscopy at particles concentration, $c$ = 0.05%, and scale bars in the insets correspond to 1 mm. The measurements reported by Nguyen et al. [26] and Li et al. [28] are also plotted in the map.

# 8 Tables

Table 1: Measured equilibrium, advancing and receding contact angles on non-heated and heated substrates for 1.1±0.2 µL water droplet containing polystyrene colloidal particles. The values of contact angle hysteresis (CAH) are also listed for all cases.

| Sr. No. | $T_s$ (°C) | $c$ (v/v, %) | Contact angles on glass (°) | | | | Contact angles on silicon (°) | | | |
|---|---|---|---|---|---|---|---|---|---|---|
| | | | $\theta_{eq}$ | $\theta_{adv}$ | $\theta_{rec}$ | CAH | $\theta_{eq}$ (°) | $\theta_{adv}$ (°) | $\theta_{rec}$ (°) | CAH |
| 1 | 27.5 | 0.05, 0.1, 1 | 34.3±1.8 | 50±2.1 | 7.5±1.6 | ~ 42.5 | 94.2±1.6 | 106±3.4 | 82±1.2 | ~ 24 |
| 2 | 60 | 0.05, 0.1, 1 | 35.7±3.8 | 52±1.8 | 10±1.2 | ~ 42 | 96.0±2.2 | 105±1.5 | 82±2.4 | ~ 23 |
| 3 | 90 | 0.05, 0.1, 1 | 38.2±3.1 | 53±1.2 | 10±2.4 | ~ 43 | 97.2±2.8 | 104±1.8 | 78±3.8 | ~ 26 |

Table 2: Various instantaneous evaporation time instances considered for the measurement of temperature difference $\Delta T$ (between the droplet contact line and apex point) to obtain the Marangoni velocities (eq 2), shown in Figure 4b and Figure 9b.

| Substrate temperature (°C) | Instantaneous evaporation time (s) | | | |
|---|---|---|---|---|
| | $t_1$ | $t_2$ | $t_3$ | $t_4$ |
| 27.5 | 3 | 24 | 192 | 300 |
| 60 | 1 | 8 | 32 | 60 |
| 90 | 0.5 | 2 | 8 | 12 |